\documentclass[11pt,preprint]{aastex}
\usepackage{natbib}
\usepackage{graphics}
\usepackage{graphicx}
\usepackage{xspace}
\usepackage{xcolor}
\usepackage[
  pdfstartview=FitH,
  linkcolor=blue,
  anchorcolor=blue,
  citecolor=blue,
  urlcolor=blue,
  colorlinks=true,
  breaklinks]{hyperref}

\shorttitle{SMA Survey of Protoplanetary Disks in NGC 2024}
\shortauthors{Mann et al.}

\received{2015}
\begin{document}

\title{Protoplanetary Disk Masses in the Young NGC 2024 Cluster}

\author{Rita K. Mann\altaffilmark{1}, 
Sean M. Andrews\altaffilmark{2}, 
Josh A. Eisner\altaffilmark{3},
Jonathan P. Williams\altaffilmark{4}, \\
Michael R. Meyer\altaffilmark{5},
James Di Francesco\altaffilmark{1,6},
John M. Carpenter\altaffilmark{7}, \&
Doug Johnstone\altaffilmark{1,6}
}
 \altaffiltext{1}{National Research Council Canada, 5071 West Saanich Road, Victoria, BC, Canada V9E 2E7} 
 \altaffiltext{2}{Harvard-Smithsonian Center for Astrophysics, 60 Garden Street, Cambridge, MA 02138, USA}
 \altaffiltext{3}{Steward Observatory, University of Arizona, 933 N Cherry Avenue, Tucson, AZ 85721}
 \altaffiltext{4}{Institute for Astronomy, University of Hawaii, 2680 Woodlawn Drive, Honolulu, HI 96822 USA}
 \altaffiltext{5}{ETH Z{\"u}rich, Institute for Astronomy, Wolfgang-Pauli-Strasse 27, 8093, Zurich, Switzerland}   
 \altaffiltext{6}{Department of Physics and Astronomy, University of Victoria, Victoria, BC, V8P 1A1, Canada } 
 \altaffiltext{7}{Department of Astronomy - California Institute of Technology, MC 249-17, Pasadena, CA 91125, USA}

\email{rita.mann@nrc-cnrc.gc.ca}

\begin{abstract}
We present the results from a Submillimeter Array survey of the 887\,$\mu$m 
continuum emission from the protoplanetary disks around 95 young stars in the 
young cluster NGC 2024.  Emission was detected from 22 infrared sources, with 
flux densities from $\sim$5 to 330\,mJy; upper limits (at 3\,$\sigma$) for the 
other 73 sources range from 3 to 24\,mJy.  For standard assumptions, the 
corresponding disk masses range from $\sim$0.003 to 0.2\,$M_\odot$, with upper 
limits at 0.002--0.01\,$M_\odot$.  The NGC 2024 sample has a slightly more 
populated tail at the high end of its disk mass distribution compared to other 
clusters, but without more information on the nature of the sample hosts it 
remains unclear if this difference is statistically significant or a 
superficial selection effect.  Unlike in the Orion Trapezium, there is no 
evidence for a disk mass dependence on the (projected) separation from the 
massive star IRS\,2b in the NGC 2024 cluster.  We suggest that this is due to 
either the cluster youth or a comparatively weaker photoionizing radiation 
field.
\end{abstract}

\keywords{circumstellar matter --- planetary systems: protoplanetary disks ---
solar system: formation --- stars: pre-main sequence}

\section{Introduction}\label{sec: intro}

The fundamental properties of circumstellar disks play a critical role in the 
formation and evolution of planets.  While detailed knowledge of disk 
properties has come from extensive studies of nearby associations like 
Taurus-Auriga and $\rho$ Ophiuchus \cite[e.g.,][]{beckwith90,osterloh,
andrews05,andrews07,andrews09,andrews10}, stars in these regions form in loose 
agglomerations and are generally unaffected by their external environment.  The 
majority of stars in the galaxy, including the Sun, formed in densely populated 
rich clusters \citep{lada03,porras,williams10}.  The high stellar density and 
ultraviolet radiation from nearby massive stars in these regions can affect 
disk properties, threatening their development and potentially limiting their 
lifespans \citep{bonnell03,johnstone98}.  Probing disk evolution in rich 
clusters is therefore crucial to our understanding of planet formation.

The Orion star-forming complex contains the nearest rich clusters with massive 
stars, and is arguably the best region for studying how disk properties are 
affected by their environment.  It is home to the clusters NGC 2024 
\citep[$\sim$0.5\,Myr;][]{meyerphd,ali98,levine06} and the Orion Nebula Cluster 
\citep[ONC, $\sim$1--2\,Myr;][]{reggiani11,dario}.  Near-infrared observations 
of the NGC 2024 cluster members have revealed 233 young stars \citep{meyerphd},
of which $\gtrsim$\,85\% exhibit an infrared excess indicative of warm dust in 
the inner regions of protoplanetary disks \citep{haisch00,haisch01}.  These 
stars are still deeply embedded in molecular cloud material \citep{barnes,
lada91}, in line with their suggested extreme youth.  The most massive star in 
the region is thought to be IRS\,2b, with a spectral type in the range of O8 to 
B2 \citep{bik,barnes}.  Although the earlier end of that range could more 
easily explain the radio continuum emission in the region, it has been 
suggested that a collection of slightly lower mass stars could together be 
responsible for the total ionizing flux \citep{meyer08}.  IRS\,2b is located 
5\arcsec\ northwest of the early B-type star IRS\,2 \citep{grasdalen74}, the 
brightest infrared and radio source in NGC 2024 \citep{barnes,rodriguez03}.

The relative youth of NGC 2024 makes it a particularly appealing region to 
probe the {\it initial} properties of disks.  Through comparisons with the 
older ONC \citep[and similar clusters, e.g.~$\sigma$ Ori;][]{williams13}, we 
can constrain key timescales for disk evolution in rich clusters.  With 
reference to pre-main sequence evolution models, the positions of the NGC 2024 
members in a color-magnitude diagram indicate very young ages: the 
\citet{baraffe98} models suggest $<$1\,Myr, and the \citet{dantona97} models 
argue for $\sim$0.5\,Myr \citep{meyerphd,ali98,levine06}.  The absolute ages of 
young stars are highly uncertain \citep[e.g.,][]{soderblom13}.  However, in a 
relative sense these are significantly younger ages than have been estimated 
for other clusters using the same technique and models \citep[e.g., 
see][]{eisner03}.  They overlap with the earliest stages of the evolution 
process for circumstellar material \citep[e.g.,][]{evans09}.  

Although the infrared excess emission found for most NGC 2024 members confirms 
the ubiquity of disks in this region, it traces only a small fraction of the 
disk material.  The low optical depths for the continuum emission at longer, 
(sub)millimeter wavelengths is required to quantitatively probe the masses of 
these disks.  While the disk population in the neighboring ONC has now been 
studied in some detail at (sub)millimeter wavelengths with interferometers 
\citep{williams05,eisner06,eisner08,mann10,mann14}, the disks in NGC 2024 are 
relatively unexplored.  \citet{eisner03} made the sole attempt at measuring the 
NGC 2024 disk mass distribution, using a 3\,mm survey of 150 targets with 
modest sensitivity (a disk mass upper limit of $\sim$0.035\,$M_\odot$).  They 
detected two massive disks (0.08 and 0.24\,$M_\odot$; significantly larger than 
seen in the ONC), and argued that image-stacking suggested that the average 
disk mass was $\sim$0.005\,$M_\odot$ (comparable to Taurus-Auriga and $\rho$ 
Ophiuchus).  

Here we present the results of a new Submillimeter Array (SMA) survey of the 
887\,$\mu$m continuum emission toward 95 young stars in the NGC 2024 cluster.  
Leveraging the steep scaling between the continuum flux and observing frequency 
($F_{\nu} \propto \nu^{2-4}$), these observations represent an order of 
magnitude improvement in sensitivity over the previous work in this region, and 
should be capable of detecting the average disk mass as suggested by 
\citet{eisner03}.  This survey represents the deepest attempt to measure the 
disk mass distribution in a very young, rich cluster, and thereby to probe how 
environment impacts basic disk properties.  The observations and their 
calibration are described in Section \ref{obs}.  The flux measurements and 
their estimated conversion to disk masses are presented in Section 
\ref{results}.  We make a comparison of the derived disk mass distribution with 
other regions, examine the dependence of disk mass on location in the cluster, 
and discuss the implications for planet formation in rich clusters in Section 
\ref{disc}.

\section{Observations} \label{obs}

\begin{figure}[t!]
\vskip -0.5in
\epsscale{1.0}
\plotone{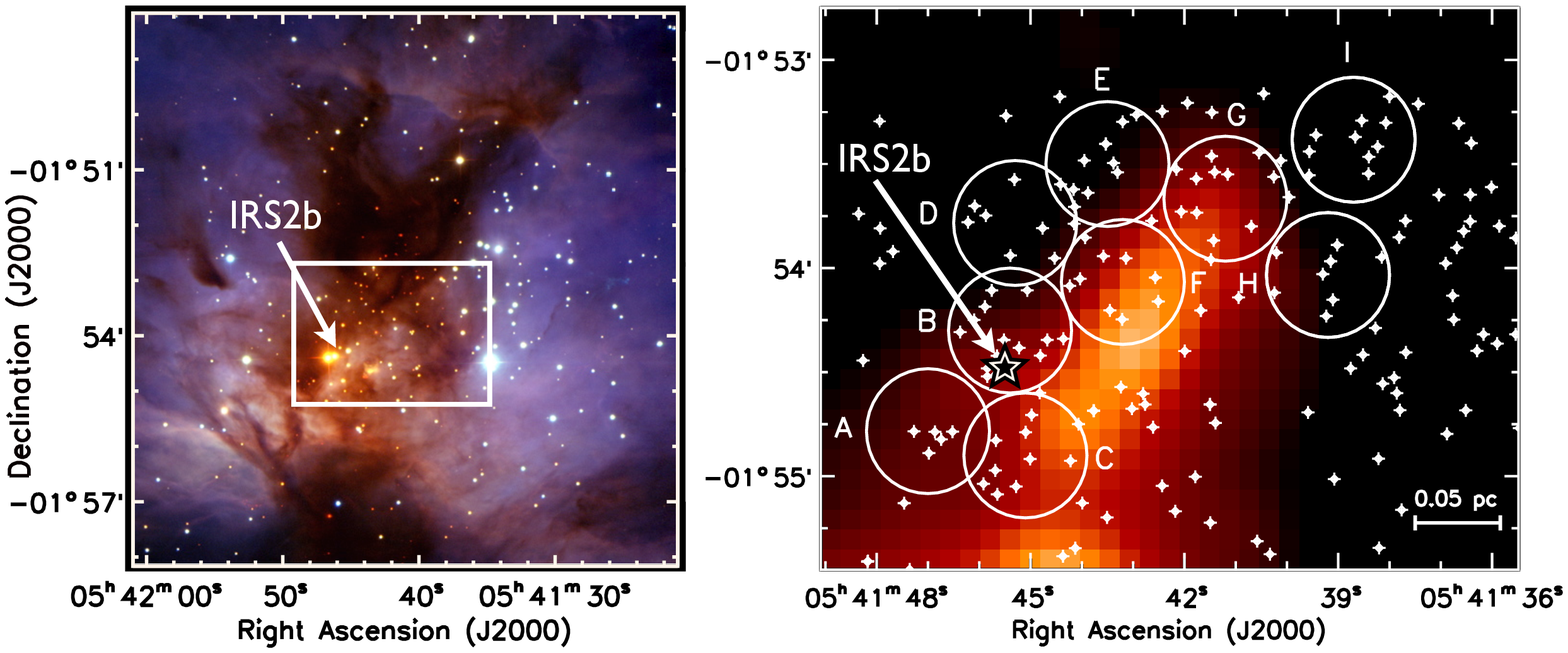}
\vskip -0.0in
\figcaption{({\it left}) A 10\,$\arcmin\times$\,10$\arcmin$ false-color
near-infrared image of NGC 2024, also known as the Flame Nebula
\citep{meyer08}.  The white box shows the region targeted in the SMA survey.
({\it right}) A JCMT-SCUBA 850\,$\mu$m image of the inset region
\citep[obtained from][]{scuba}, with the pointing locations and dimensions of
the SMA primary beam overlaid as white circles, labeled as in Table
\ref{table-obs}.  Crosses show the location of young stars identified in
$K$-band imaging \citep{meyerphd}.  The most massive star of the cluster,
IRS\,2b, is labeled in both panels. \label{pointings}}
\end{figure}

Observations of nine distinct pointings containing a total of 95 young stars 
were conducted with the SMA \citep{ho} in the fall of 2011, using the compact 
array configuration (baselines of $\sim$8--50\,m).  The pointing centers are 
listed in Table \ref{table-obs} and shown in Figure \ref{pointings}, and were 
chosen to maximize the number of young stars imaged in each (35\arcsec\ FWHM) 
primary beam while minimizing contamination from the bright, non-uniform 
molecular cloud background in the region (see Section \ref{results}).  The SMA 
double sideband receivers were tuned to a local oscillator (LO) frequency of 
338.213\,GHz (887\,$\micron$, see Table 1).  Each sideband provided 4\,GHz of 
bandwidth, centered $\pm$\,5 GHz from the LO frequency.  Three observing tracks 
were shared between three separate pointings in each.  The observations of the 
NGC 2024 fields were interleaved with nearby gain calibrators on 15\,minute 
intervals.  Weather conditions for all observations were good, with $<$2\,mm of
precipitable water vapor, $\tau$(225 GHz)\,$<$\,0.1.  Table \ref{table-obs} 
summarizes the relevant observational information.

\begin{figure}[t!]
\epsscale{1.0}
\plotone{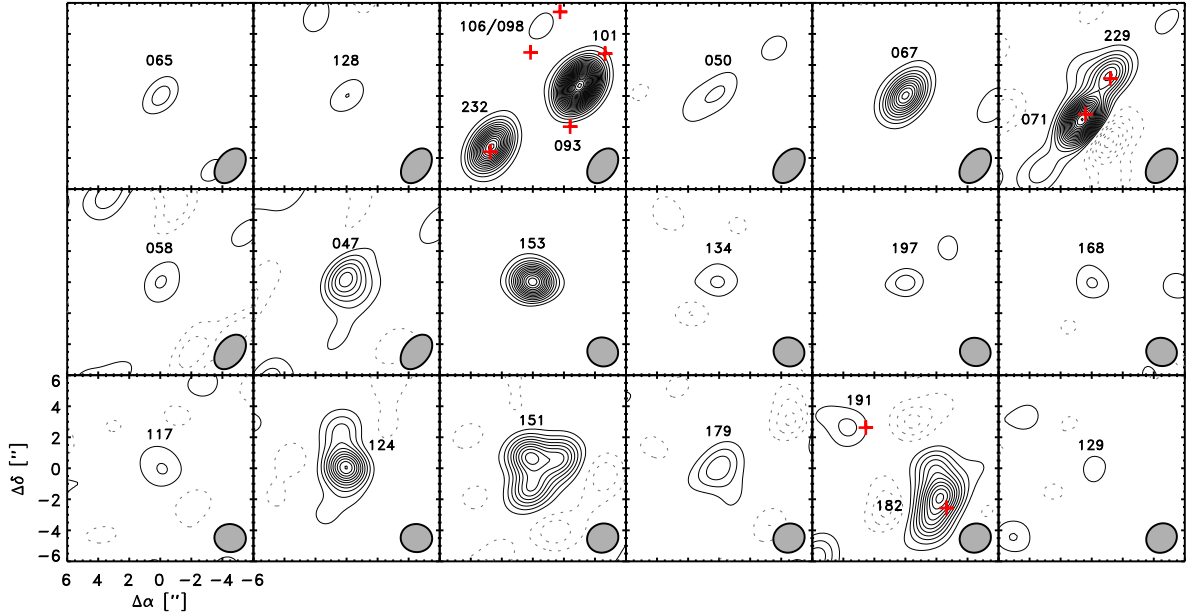}
\figcaption{Synthesized images of the 887\,$\mu$m continuum emission toward the
22 detected infrared sources (see Table \ref{table-det}).  Each panel is 
12\arcsec\ ($\sim$5000\,AU) on a side, and includes labels in the 
\citet{meyerphd} IRC designation.  Contours are drawn at 3\,$\sigma$ intervals, 
and the synthesized beam dimensions are shown in the bottom right corners of 
each panel.  In all but three panels, the image center corresponds to the 
infrared source position; otherwise, those positions are marked with crosses.  
The third panel includes the two cases with multiple possible identifications; 
see Table \ref{table-det}.
\label{cont}}
\end{figure}

The raw visibilities were calibrated using the {\tt MIR} software package.  
Passband calibration was conducted using the bright, compact radio sources, 3C 
279, 3C 84, or J0854+201.  The absolute flux scale was derived from 
observations of Titan and Uranus (and checked against 3C 111), and is accurate 
to $\sim$10\%.  Amplitude and phase calibration were performed using 
observations of the nearby sources J0423$-$013, J0530+135, and J0607$-$085.  
The calibrated visibilities were naturally weighted and Fourier inverted, then 
{\tt CLEAN}ed to generate the synthesized continuum maps shown in Figure 
\ref{cont} using {\tt MIRIAD} \citep{sault}.  The synthesized maps were created 
after eliminating projected antenna spacings shorter than 27\,k$\lambda$, to 
filter out extended emission on size scales $\geq$\,7.5$\arcsec$.  This scale 
was chosen to preserve the compact emission from the disks, while minimizing 
contamination from the bright, extended cloud background; it is the same as 
used in the analysis of dust emission from disks in the ONC \citep{mann10} for 
the sake of consistency.  Simulations of the background (see Section 
\ref{results}) confirmed that the 27\,k$\lambda$ cutoff resolves out most of 
the extended emission and thereby helps reduce the effective RMS noise levels 
by roughly a factor of $\sim$2.

\section{Results}\label{results}

Continuum emission was detected toward 22 near-infrared sources at $>$3$\times$ 
the measured RMS noise level (see Table \ref{table-det}).\footnote{An 
additional $\sim$20 emission peaks (each at $\sim$3\,$\sigma$) were identified, 
but are not coincident with any known near-infrared sources (some of them are 
visible in Fig.~\ref{cont}).  This is the expected number of 3\,$\sigma$ noise 
peaks over such a large survey area (perhaps even an under-estimate when 
considering the sparse Fourier sampling), although we cannot rule out the 
possibility that some are real sources associated with deeply embedded 
objects.}  The observed flux density, $F_{\rm obs}$, for each source was 
determined with a Gaussian fit in the image plane, after correcting for primary 
beam attenuation.  The detected sources were associated with their stellar 
counterparts by reference to the near-infrared catalog of \citet{meyerphd}.  In 
two cases, IRC\,106/098 and IRC\,101/093 (see Table \ref{table-det}), it is 
difficult to unambiguously distinguish two potential associations with infrared 
sources.  For the former, observations with improved sensitivity and resolution 
(and ideally Fourier sampling) will be required to robustly differentiate the 
options.  On the other hand, it is entirely possible that the very bright 
emission near IRC\,101/093 is unassociated with either source, but instead is 
tracing an embedded source at an earlier evolutionary stage.  Another 73 
infrared sources were covered in the survey, but not detected.  Limits on their 
submillimeter continuum flux densities were determined from local measurements 
of the RMS noise level, and are also listed in Table \ref{table-det}.  

The observed flux densities are the linear combination of several emission 
contributors; free-free radiation from ionized material ($F_{\rm ff}$), and 
thermal radiation from dust in the surrounding molecular cloud ($F_{\rm 
cloud}$) and the disk ($F_{\rm disk}$).  We assumed that any free-free emission 
is optically thin, with a spectrum $F_{\rm ff} \propto \nu^{-0.1}$, and used 
deep VLA 3.6\,cm measurements \citep[][see Table \ref{table-det}]{rodriguez03} 
when available as normalizations to extrapolate $F_{\rm ff}$ up to 
887\,$\mu$m.  Only four of the VLA 3.6\,cm sources overlap with the SMA 
887\,$\mu$m detections.  Upper limits at 3.6\,cm are sufficiently low 
($\sim$50\,$\mu$Jy at 3\,$\sigma$) that extrapolated estimates of free-free 
contamination for the other targets are considered negligible.  The observed 
targets are embedded in their host cloud, which itself produces significant 
dust emission on large spatial scales.  To estimate that cloud contribution at 
each location in the SMA maps, we simulate the SMA response to its large-scale 
emission as observed with the SCUBA instrument on the James Clerk Maxwell 
Telescope \citep[JCMT;][]{scuba}, following the approach of \citet{mann10}.  
For each SMA pointing, we Fourier transformed the appropriate JCMT map, sampled 
the emission onto the observed $u,v$-tracks, and generated {\tt CLEAN}ed maps.  
The cloud emission, $F_{\rm cloud}$, was then estimated toward each source 
location.  We found that cloud contamination was typically $<$10\%, 
considerably lower than in the ONC.  After accounting for any free-free or 
cloud contamination, the remainder of $F_{\rm obs}$ was associated with the 
disk, $F_{\rm disk}$.  Table \ref{table-det} lists the decomposed contributions 
for each source.  

Relatively little is known about the nature of the infrared sources associated 
with submillimeter emission in this survey.  Given their neutral or modestly 
red $JHK$ colors, we are making the assumption that the vast majority of them 
are in the so-called Class II or T Tauri stage, young stars with disks but no 
remnant envelope material.  It is possible that the few most luminous sources 
could be in the earlier Class I stage, if their observed near-infrared emission 
is primarily tracing scattered light from outflow cavities in their envelope 
structures \citep{eisner12,sheehan14}.  As we noted above, the very brightest 
source in this sample cannot be unambiguously associated with an infrared 
source; it is plausible that the emission may originate in a dense envelope 
around a Class 0 protostar.  Without a more complete set of ancillary 
information (e.g., full spectral energy distributions), a refined 
classification is not yet feasible.

Following \citet{eisner03}, we aimed to constrain the mean emission level of 
the 73 undetected sources in the SMA fields by ``stacking" the data.  However, 
there is sufficient concern with combining data in the image plane when the 
individual fields were sampled so sparsely in the Fourier domain.  Instead we 
performed a complementary analysis on the visibilities.  First, we removed the 
detected sources by subtracting Gaussian models of the emission from the 
observed visibilities.  We then generated 73 permutations of the visibility 
data, each with a phase shift that accounts for the location of the undetected 
cluster member.\footnote{Technically this includes substantial duplication of 
the data.  However, the individual sources are located far enough apart that 
their mutual contributions at any given phase shift are considered 
negligible.}  Those permutations were co-added and then imaged as described in 
Section \ref{obs}.  We found no emission associated with this stacked dataset, 
and placed a 3\,$\sigma$ upper limit of $\sim$2.4\,mJy on the ensemble 
average.  Assuming a typical spectrum that scales like $F_{\nu} \propto 
\nu^{2-3}$, this limit is $\sim$4--10$\times$ lower than the ensemble mean flux 
density that was estimated from image-plane stacking by \citet{eisner03}.  

We assume that the observed continuum emission is (mostly) optically thin, and 
therefore a sensitive probe of the dust mass.  Since most of the emission 
originates in the cool, nearly isothermal outer regions of a disk, we estimate 
the mass as
\begin{equation}\label{masseqn}
M_{\rm disk} = \frac{d^2 \, F_{\rm disk}}{\kappa_{\nu} \, B_{\nu}(T)},
\end{equation}
where $d$ is the distance, $\kappa_{\nu}$ is the opacity per gram of the disk 
material, and $B_{\nu}(T)$ is the Planck function at a characteristic 
temperature \citep[e.g.,][]{beckwith90}.  Disk masses were calculated for the 
22 detected sources (see Table \ref{table-det}) using Eq.~\ref{masseqn} and 
standard assumptions (for ease of comparison with other studies): a 
characteristic dust temperature $T = 20$\,K, the \citet{beckwith90} opacity 
$\kappa_{\nu} = 0.034$\,cm$^2$\,g$^{-1}$ at 887\,$\mu$m (which implicitly 
assumes a 100:1 gas-to-dust mass ratio), and a distance $d = 415$\,pc to NGC 
2024, based on observations of B-type stars in the cluster 
\citep{Anthony-twarog}.  Disk emission was inferred toward the source IRS 2 
(IRC 232), an early B-type star; in calculating its mass, we adopted a higher 
dust temperature of 40\,K \citep[e.g., see][]{beuther02,Sridharan02}.
Systematic uncertainties in the $M_{\rm disk}$ estimates are dominated by the 
poorly constrained values of the dust opacities, which are ambiguous at the 
order of magnitude level \citep[e.g.,][]{henning96}.  Some relatively minor 
additional uncertainties from optical depth effects could also contribute, 
especially given the unknown disk structures: \citet{andrews05} estimated that 
these are on the order of $\sim$10\%\ for the fainter luminosities that 
characterize this sample, but could rise to as high as $\sim$50\%\ at the high 
luminosity end.  

The overall disk mass sensitivity of this SMA survey depends on the locations 
of each target relative to the field (pointing) center, the varying levels of 
cloud emission, and any free-free emission contributions.  A mass completeness 
level for the survey was estimated based on Monte Carlo simulations.  Synthetic 
disks (point sources) with emission appropriate for a given $M_{\rm disk}$ (see 
Eq.~1) were injected into the large-scale emission maps from the JCMT, Fourier 
inverted and sampled onto the observed spatial frequencies, and then each field 
was imaged as in Section 2.  By measuring the fraction of synthetic targets 
that were detected ($>$3\,$\sigma$) for each input $M_{\rm disk}$ in these 
Monte Carlo simulations, we determined that the survey is essentially 100\%\ 
complete for $M_{\rm disk} \geq 0.01$\,$M_{\odot}$, and roughly 50\%\ complete 
for $M_{\rm disk} \geq 0.004$\,$M_{\odot}$.  The upper limit on the stacked 
ensemble of undetected sources described above corresponds to 
$\sim$0.001\,$M_{\odot}$.

Four of the 13 sources detected by \citet{eisner03} at 3\,mm were also 
detected in the SMA survey at 887\,$\mu$m: IRC\,124, IRC\,071, IRC\,101, and 
IRS\,2 (their sources 1, 4, 8, and 9, respectively).  The remaining nine 
sources were either outside our survey area (their sources 3, 11, 12, and 13) 
or were not detected with the SMA (IDs 2, 5, 6, 7, 9, and 10); the latter cases 
likely indicate that the radiation detected at 3\,mm was generated by a 
non-dust emission mechanism.  The observed flux ratios can be used to constrain 
the spectral slopes between 887\,$\mu$m and 3\,mm of the targets with 
overlapping detections.  We find spectral indices, $\alpha$ where $F_{\nu} 
\propto \nu^{\alpha}$, of $2.8\pm0.2$ (IRC\,124 = source 1), $2.8\pm0.1$ 
(IRC\,071 = source 4), $2.6\pm0.1$ (IRC\,101 = source 8), and $0.1\pm0.1$ 
(IRS\,2 = source 9); the IRS\,2 measurement is obviously strongly impacted by 
free-free contamination at 3\,mm.  These spectral indices are at the higher end 
of the distributions inferred in $\sim$1--3\,Myr-old star-forming clusters 
\citep{ricci10,ricci10b,ricci11,ricci11b}, perhaps hinting at some age 
evolution in the disk-integrated opacity spectrum due to dust grain growth 
\citep[e.g.,][]{miyake93,draine06}.  More measurements for young sources in NGC 
2024 would be desireable.    

\begin{figure}[t!]
\epsscale{0.6}
\plotone{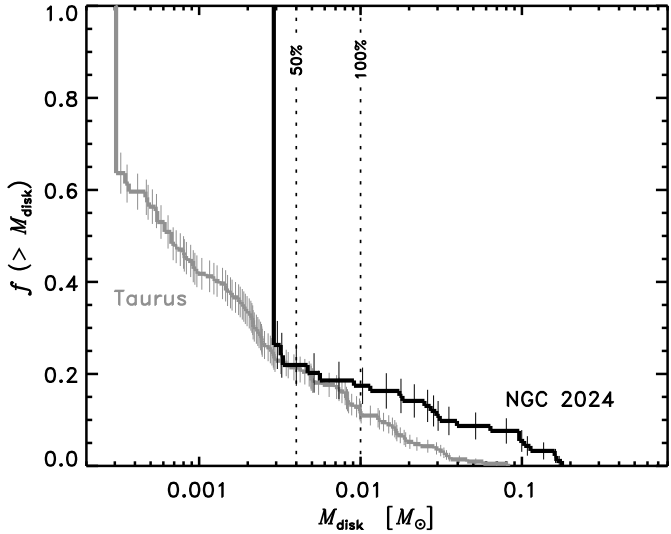}
\figcaption{The cumulative distribution of disk masses in the NGC 2024 survey,
constructed with the Kaplan-Meier product limit estimator to account for upper
limits.  The estimated 50 and 100\%\ completeness levels are marked as dotted
vertical lines.  The analogous distribution for the Taurus star-forming region
\citep{andrews13} is shown in gray for reference.  The upper end of the NGC
2024 disk mass distribution favors slightly higher masses, as might be
expected for a younger cluster, although the selection effects of this sample
are not yet well characterized.  \label{cdf}}
\end{figure}

\section{Discussion}\label{disc}

We conducted a large-scale survey of the 887\,$\mu$m continuum emission toward 
95 young stars in the $\sim$0.3\,Myr-old NGC 2024 cluster using the SMA.  
Assuming standard conversions for optically thin dust emission, this survey is 
complete down to a disk mass limit of $\sim$0.01\,$M_{\odot}$ (3\,$\sigma$), 
although in some regions the survey is slightly more sensitive.  We detected a 
total of 22 disks ($23\pm5$\%\ of the sample), including four that were 
previously detected at 3\,mm by \citet{eisner03}.  

Figure \ref{cdf} shows the cumulative distribution of disk masses in this 
survey, where we have incorporated the upper limits by employing the 
Kaplan-Meier product limit estimator for a censored sample \citep{feigelson}.  
Although these data are sensitivity-limited to probe only relatively massive 
disks, we find that the fraction of disks with large $M_{\rm disk}$ is 
relatively high: $\sim$20\%\ have $M_{\rm disk} > 0.01$\,$M_{\odot}$, and 
$\sim$10\%\ have $M_{\rm disk} > 0.1$\,$M_{\odot}$.  Taken at face value, this 
suggests that the high-mass tail of the NGC 2024 $M_{\rm disk}$ distribution is 
more populated than in slightly older ($\sim$1--3\,Myr) clusters like the ONC 
\citep{mann14}, Ophiuchus \citep{andrews07}, and Taurus \citep[see the 
corresponding distribution function in Fig.~\ref{cdf} for a direct 
comparison;][]{andrews05,andrews13}.  The standard censored two-sample tests 
advocated by \citet{feigelson} indicate a marginal ($\sim$2\,$\sigma$) 
quantitative offset, with the NGC 2024 distribution shifted to higher masses by 
a factor of $\sim$1.5--2.

However, such comparisons can be misleading if they do not account for 
selection biases.  Disk masses are known to depend on factors like the (host) 
mass \citep{andrews13}, multiplicity \citep{harris12,akeson14}, and 
evolutionary state \citep{andrews05} of the target.  Unfortunately, little is 
known about these properties for the NGC 2024 sample.  If we make the 
assumptions that this survey has targets drawn from the same host mass function 
and with the same multiplicity statistics as in Taurus, and suggest that all of 
the near-infrared sources in the SMA fields harbor disks with no envelopes, we 
can use the Monte Carlo approach advocated by \citet{andrews13} to compare with 
the reference $M_{\rm disk}$ distribution in Taurus.  The results indicate that 
the two $M_{\rm disk}$ distributions are statistically indistinguishable (given 
the current data), suggesting that there is relatively little evolution in 
$M_{\rm disk}$ up to a few Myr.  More robust constraints on changes in this 
distribution would require a better characterization of the NGC 2024 targets, 
and secondarily an expanded continuum census.

\begin{figure}[ht!]
\epsscale{0.6}
\plotone{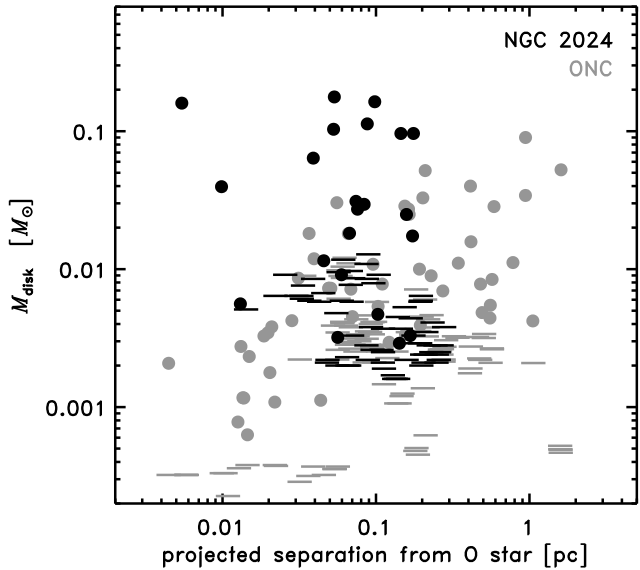}
\figcaption{Disk masses in the NGC 2024 ({\it black}) and ONC ({\it gray}; see 
\citealt{mann10,mann14}) clusters as a function of their projected separations 
from the nearest massive star, IRS\,2b and $\theta^1$\,Ori C, respectively.  
Circles represent submillimeter continuum detections of dust disk emission, and 
horizontal line segments mark 3\,$\sigma$ upper limits.  The depletion in $M_{\rm 
disk}$ at small projected separations seen for the ONC is not apparent for the 
NGC 2024 cluster, perhaps because it represents an earlier evolutionary stage 
or due to the (presumably) comparatively weaker photoionizing radiation field 
present.  \label{massdist}}
\end{figure}

But independent of these (potential) selection effects, a particularly 
interesting comparison can still be made between the disks in the NGC 2024 
cluster and the ONC, since both regions host high-mass stars in their immediate 
environments that could potentially modify the disk mass distribution.  Sources 
in the ONC found within $\sim$0.03\,pc of the massive star $\theta^1$\,Ori C 
have systematically lower $M_{\rm disk}$ than those at larger separations 
\citep{mann09b,mann10,mann14}, reflecting the consequences of external 
evaporation from strong photoionizing sources on disk dissipation timescales 
\citep{johnstone98,storzer,richling00,scally01,matsuyama,adams04}.  Figure 
\ref{massdist} shows $M_{\rm disk}$ as a function of the projected distance 
from the most massive stars in both the NGC 2024 cluster and the ONC, IRS\,2b 
and $\theta^1$\,Ori C, respectively.  Unlike the ONC, we find no evidence of a 
distance-dependent disk mass distribution in NGC 2024.  Although the total 
number of disk detections in the NGC 2024 region is limited, several massive 
disks identified here are located $<$0.01\,pc from IRS\,2b.

We consider two likely, and not mutually exclusive, reasons for this difference 
between NGC 2024 and the ONC.  First is an evolutionary argument, based on the 
fact that NGC 2024 appears to be considerably younger than the ONC 
\citep[see][]{meyerphd,eisner03}.  That relative youth could mean that we are 
observing more of a primordial $M_{\rm disk}$ distribution in NGC 2024, before 
external evolutionary processes like photoevaporation have had time to make a 
significant impact.  Second is an environmental distinction, in that IRS\,2b is 
an intrinsically less luminous source than $\theta^1$\,Ori C, and therefore 
could produce a substantially weaker photoionizing radiation field that is less 
capable of stripping material from its surrounding disks.  \citet{bik} 
suggested that IRS\,2b has a spectral type of O8--B2, although the radio 
continuum flux measured by \citet{barnes} indicates that the earlier type is 
more appropriate.  However, \citet{meyer97} instead suggested that early 
B-types are the preferred spectroscopic classification for IRS\,2b, and that it 
may not be the sole or even dominant source of ionizing radiation in the region 
\citep[see also][]{meyer08}.  A small group of weaker ionizing sources might 
provide sufficiently attenuated mass-loss rates and explain the absence of a 
separation dependence on the $M_{\rm disk}$ distribution.  Even if IRS\,2b was 
more luminous than described, NGC 2024 is more heavily extincted than the ONC, 
and the high energy photons could be easily absorbed by the cloud.

Future, more sensitive observations of NGC 2024 with the Atacama Large 
Millimeter Array (ALMA) will permit an exploration of the full disk mass 
distribution in this young region, facilitating stronger constraints on disk 
dissipation by ultraviolet photoevaporation, on the evolutionary timescales of 
disks in rich clusters, and on the initial conditions of the planet formation 
process for the majority of stars in the galaxy.  Such campaigns should 
necessarily be coupled with a more comprehensive characterization of the NGC 
2024 stellar population.

\acknowledgments We thank the referee for a very helpful review.  The 
Submillimeter
Array is a joint project between the Submillimeter Astrophysical Observatory
and the Academica Sinica Institute of Astronomy and Astrophysics and is funded
by the Smithsonian Institution and the Academica Sinica.

\clearpage
\begin{deluxetable}{cccccccc}
\tablecolumns{8}
\tablewidth{0pc}
\tablecaption{Summary of Submillimeter Array Observations\label{table-obs}}
\tablehead{  \colhead{Field} & \colhead{$\alpha$ (J2000)} &
             \colhead{$\delta$ (J2000)} & 
             \colhead{UT Date} &
             \colhead{$\tau$} & \colhead{$\sigma$ (mJy/beam)} &
             \colhead{$\theta_b$ (\arcsec)} & \colhead{PA$_b$ (\degr) }  \\
        \colhead{(1)} & \colhead{(2)} & \colhead{(3)} & \colhead{(4)} &
        \colhead{(5)} & \colhead{(6)} & \colhead{(7)} & \colhead{(8)}
}
\startdata
A & 05 41 47.0 & -01 54 47 & 2011 Sep 23 & 0.06-0.07 & 1.0 & 2.5 x 1.5 & 158 \\
B & 05 41 45.4 & -01 54 18 & 2011 Sep 23 & 0.06-0.07 & 3.0 & 2.5 x 1.5 & 158 \\
C & 05 41 45.1 & -01 54 54 & 2011 Sep 23 & 0.06-0.07 & 3.5 & 2.5 x 1.5 & 158 \\
D & 05 41 45.3 & -01 53 47 & 2011 Oct 18 & 0.03-0.07 & 1.4 & 1.8 x 1.5 &  15 \\
E & 05 41 43.5 & -01 53 30 & 2011 Oct 18 & 0.03-0.07 & 1.0 & 1.8 x 1.5 &  15 \\
F & 05 41 43.2 & -01 54 04 & 2011 Oct 18 & 0.03-0.07 & 4.6 & 1.8 x 1.5 &  15 \\
G & 05 41 41.2 & -01 53 40 & 2011 Dec 29 & 0.08-0.15 & 2.2 & 2.2 x 1.7 &   5 \\
H & 05 41 39.2 & -01 54 02 & 2011 Dec 29 & 0.08-0.15 & 1.0 & 2.2 x 1.7 &   5 \\
I & 05 41 38.7 & -01 53 23 & 2011 Dec 29 & 0.08-0.15 & 1.4 & 2.2 x 1.7 &   5 \\
\enddata
\tablecomments{All observations were conducted at 887\,$\mu$m using the compact 
array configuration.  Col.~(1): SMA pointing, as labeled in Figure 
\ref{pointings}.  Cols.~(2, 3): Phase center coordinates.  Col.~(4): UT Date of 
observation.  Col.~(6): Range of zenith optical depths at 225\,GHz.  Col.~(8): 
RMS noise level measured in emission-free regions within the primary beam of 
the naturally-weighted synthesized maps.  Cols.~(9, 10): Dimensions and 
orientations of the synthesized beams.}
\end{deluxetable}

\clearpage
\vskip 0in
\begin{deluxetable}{ccllrccrr}
\tablecolumns{9}
\tablewidth{0pc}
\tablecaption{Inferred Disk Fluxes and Masses\label{table-det}}
\tablehead{  \colhead{Source} & \colhead{Field} &
             \colhead{$\alpha$ (J2000)} & \colhead{$\delta$ (J2000)} &
             \colhead{$F_{\rm obs}$} &
             \colhead{$F_{\rm ff}$} & \colhead{$F_{\rm cloud}$} &
             \colhead{$F_{\rm disk}$} & \colhead{$M_{\rm disk}$} \\
             
        \colhead{(IRC)} & \colhead{} & \colhead{} & \colhead{} &
        \colhead{(mJy)} & \colhead{(mJy)} &
        \colhead{(mJy)} & \colhead{(mJy)} & 
        \colhead{(0.01\,$M_{\odot}$)}  \\
        
        \colhead{(1)} & \colhead{(2)} & \colhead{(3)} & \colhead{(4)} &
        \colhead{(5)} & \colhead{(6)} & \colhead{(7)} & \colhead{(8)} & 
        \colhead{(9)}  
}
\startdata
065      & A & 05 41 46.9 & -01 54 46.9 & $7.2\pm1.2$   & 0.8  &  0.5 & 5.9   & 0.3$\pm$0.1  \\
128      & B & 05 41 45.1 & -01 54 06.9 & $21.2\pm3.0$  &     &  0   & 21.8  & 1.2$\pm$0.2  \\
106/098  & B & 05 41 45.6 & -01 54 22.4 & $10.7\pm3.0$  &     &  0.3 & 10.5  & 0.6$\pm$0.2  \\
101/093* & B & 05 41 45.4 & -01 54 26.3 & $298.6\pm3.0$ & 0.7  &  0.3 & 297.7 & 16.0$\pm$0.2 \\
232* (IRS\,2)     & B & 05 41 45.8 & -01 54 30.2 & $209.4\pm3.0$ & 12.0 &  0   & 199.3 & 4.0$\pm$0.2  \\
050      & C & 05 41 45.8 & -01 55 06.2 & $50.7\pm3.0$  &     &  0.1 & 50.6  & 2.7$\pm$0.3  \\
067      & C & 05 41 45.1 & -01 54 47.3 & $119.0\pm3.0$ &     &  0.2 & 118.9 & 6.4$\pm$0.2  \\
071*     & C & 05 41 44.1 & -01 54 45.8 & $330.5\pm3.0$ & 0.2  &  0.1 & 330.1 & 17.7$\pm$0.3 \\
229      & C & 05 41 44.0 & -01 54 43.1 & $192.6\pm3.0$ &     &  0.2 & 192.5 & 10.3$\pm$0.4 \\
058      & C & 05 41 44.1 & -01 54 54.6 & $34.1\pm3.0$  &     &  0.1 & 34.0  & 1.8$\pm$0.3  \\
047      & C & 05 41 44.1 & -01 55 06.5 & $210.7\pm3.0$ &    &  0.1 & 210.6 & 11.3$\pm$0.4 \\
153      & D & 05 41 44.7 & -01 53 48.7 & $55.0\pm1.2$  &     &  0   & 55.2  & 3.0$\pm$0.1  \\
134      & D & 05 41 44.7 & -01 54 01.6 & $17.1\pm1.2$  &     &  0.1 & 17.0  & 0.9$\pm$0.1  \\
197      & E & 05 41 43.5 & -01 53 24.8 & $6.1\pm1.0$   &     &  0.8 & 5.3   & 0.3$\pm$0.1  \\
168      & E & 05 41 44.1 & -01 53 42.3 & $9.7\pm1.0$   &     &  0.9 & 8.8   & 0.5$\pm$0.1  \\
117      & F & 05 41 43.2 & -01 54 15.6 & $58.1\pm4.5$  &     &  0.5 & 57.7  & 3.1$\pm$0.3  \\
124*     & F & 05 41 42.5 & -01 54 08.8 & $305.5\pm4.5$ &     &  0.9 & 304.6 & 16.4$\pm$0.3 \\
151      & G & 05 41 41.5 & -01 53 48.6 & $179.5\pm2.0$ &     &  0.1 & 179.4 & 9.6$\pm$0.1  \\
179      & G & 05 41 41.7 & -01 53 35.2 & $46.3\pm2.0$  &     &  0.0 & 46.4  & 2.5$\pm$0.1  \\
191      & G & 05 41 41.5 & -01 53 26.7 & $32.3\pm2.0$  &    &  0   & 33.1  & 1.8$\pm$0.2  \\
182      & G & 05 41 41.1 & -01 53 31.2 & $172.3\pm2.0$ &     &  0   & 172.7 & 9.3$\pm$0.1  \\
129      & H & 05 41 40.2 & -01 54 06.6 & $6.3\pm1.3$   &     &  0   & 6.3   & 0.3$\pm$0.1  \\
\hline
Non Detections &  &  &  & 3$\sigma$ Upp Limit &  &  &  &  \\
\hline
066      & A & 05 41 46.5   & -01 54 47.2     & $4.0$     & 0.8  &  0.1    & 3.1       & 0.2  \\
068      & A & 05 41 46.7   & -01 54 45.1     & $3.8$      &      &  0.2    & 3.6       & 0.2  \\
062      & A & 05 41 46.8   & -01 54 49.3     & $3.8$      &      &  0.2    & 3.6       & 0.2  \\
060      & A & 05 41 47.0   & -01 54 53.1     & $3.9$      &      &  0.3    & 3.6       & 0.2  \\
064      & A & 05 41 47.3   & -01 54 47.1     & $3.7$      &      &  0.1    & 3.6       & 0.2  \\
107      & B & 05 41 44.7   & -01 54 20.7     & $12.0$    &      &  0.2    &11.8      & 0.6  \\
231      & B & 05 41 44.7   & -01 54 31.4     & $17.1$    &      &  0.2    &16.9      & 0.9  \\
099      & B & 05 41 44.8   & -01 54 25.4     & $12.0$    & 0.2  &  0.1    &11.8      & 0.6  \\
134      & B & 05 41 44.9   & -01 54 01.8     & $17.8$    &      &  0.1    &17.7      & 1.0  \\
102      & B & 05 41 45.2   & -01 54 23.1     & $9.7$      &      &  0.2    & 9.5       & 0.5  \\
127      & B & 05 41 45.7   & -01 54 06.5     & $12.7$    &      &  0.1    &12.6      & 0.7  \\
122      & B & 05 41 45.9   & -01 54 11.3     & $11.0$    & 0.1  &  0.2    &10.7      & 0.6  \\
114      & B & 05 41 46.1   & -01 54 15.0     & $11.5$    &      &  0.1    &11.4      & 0.6  \\
109      & B & 05 41 46.4   & -01 54 18.4     & $14.2$    &      &  0.1    &14.0      & 0.8  \\
073      & C & 05 41 44.9   & -01 54 42.6     & $12.0$    &      &  0.1    &11.9      & 0.6  \\
059      & C & 05 41 45.0   & -01 54 55.1     & $9.1$      & 0.2  &  0.2    & 8.7       & 0.5  \\
052      & C & 05 41 45.3   & -01 55 03.0     & $10.9$    &      &  0.1    &10.8      & 0.6  \\
063      & C & 05 41 45.7   & -01 54 49.7     & $10.9$    &      &  0.2    &10.7      & 0.6  \\
057      & C & 05 41 45.7   & -01 54 58.3     & $11.0$    & 0.2  &  0.1    &10.6      & 0.6  \\
053      & C & 05 41 45.9   & -01 55 02.2     & $11.3$    &      &  0.2    &14.2      & 0.8  \\
157      & D & 05 41 44.1   & -01 53 47.5     & $ 7.1$    &      &  0       & 7.1      & 0.4  \\
175      & D & 05 41 44.4   & -01 53 36.2     & $ 7.0$    &      &  0.1    & 6.9      & 0.4  \\
138      & D & 05 41 44.5   & -01 53 57.4     & $ 6.2$    &      &  0       & 6.2      & 0.3  \\
176      & D & 05 41 45.3   & -01 53 34.8     & $ 4.9$    &      &  0.2    & 4.7      & 0.3  \\
139      & D & 05 41 45.4   & -01 53 56.5     & $ 4.4$    &      &  0       & 4.4      & 0.2  \\
161      & D & 05 41 45.9   & -01 53 45.1     & $ 4.2$    &      &  0.1    & 4.1      & 0.2  \\
167      & D & 05 41 46.1   & -01 53 42.3     & $ 5.0$    &      &  0.1    & 4.9      & 0.3  \\
156      & D & 05 41 46.2   & -01 53 47.0     & $ 5.4$    &      &  0.1    & 5.3      & 0.3  \\
206      & E & 05 41 42.9   & -01 53 16.3     & $ 5.4$    &      &  0.1    & 5.3      & 0.3  \\
202      & E & 05 41 43.2   & -01 53 18.4     & $ 4.3$    &      &  0.2    & 4.1      & 0.2  \\
181      & E & 05 41 43.3   & -01 53 32.8     & $ 3.1$    &      &  0.2    & 2.9      & 0.2  \\
187      & E & 05 41 43.3   & -01 53 30.2     & $ 3.0$    &      &  0.2    & 2.8      & 0.2  \\
193      & E & 05 41 43.5   & -01 53 26.4     & $ 3.1$    &      &  0.1    & 3.0      & 0.2  \\
170      & E & 05 41 43.8   & -01 53 38.6     & $ 3.7$    &      &  0.1    & 3.6      & 0.2  \\
188      & E & 05 41 43.9   & -01 53 29.3     & $ 3.3$    &      &  0.1    & 3.2      & 0.2  \\
172      & E & 05 41 44.1   & -01 53 37.6     & $ 4.1$    &      &  0       & 4.1      & 0.2  \\
175      & E & 05 41 44.4   & -01 53 36.2     & $ 4.7$    &      &  0.1    & 4.6      & 0.2  \\
133      & F & 05 41 42.5   & -01 54 03.1     & $17.1$    &     &  0.2    &16.9      & 0.9  \\
115      & F & 05 41 42.8   & -01 54 16.1     & $20.4$    &      &  0.1    &20.3      & 1.0  \\
140      & F & 05 41 43.1   & -01 53 57.5     & $14.9$    &      &  0.3    &14.6      & 0.8  \\
121      & F & 05 41 43.4   & -01 54 12.5     & $16.1$    &      &  0       &16.1      & 0.9  \\
143      & F & 05 41 43.5   & -01 53 56.9     & $15.9$    & 0.1  &  0.2    &15.6      & 0.8  \\
150      & F & 05 41 43.9   & -01 53 51.5     & $23.9$    &      &  0.2    &23.7      & 1.3  \\
132      & F & 05 41 44.0   & -01 54 03.2     & $18.3$    &      &  0.3    &18.0      & 1.0  \\
131      & F & 05 41 44.2   & -01 54 21.9     & $21.9$    &      &  0.3    &21.6      & 1.2  \\
178      & G & 05 41 40.0   & -01 53 35.7     & $ 11.9$    &      &  0.1    & 11.8      & 0.6  \\
180      & G & 05 41 40.2   & -01 53 34.3     & $ 10.8$    &      &  0.1    & 10.7      & 0.6 \\
194      & G & 05 41 40.5   & -01 53 27.4     & $ 11.0$    &      &  0       & 11.0      & 0.6  \\
158      & G & 05 41 40.6   & -01 53 48.5     & $  8.3$     &      &  0       &  8.3      &  0.4  \\
165      & G & 05 41 41.7   & -01 53 44.6     & $  7.0$     &      &  0       &  7.0      &  0.4 \\
166      & G & 05 41 42.0   & -01 53 44.3     & $  8.4$     &      &  0.1    &  8.3      &  0.4 \\
185      & G & 05 41 42.1   & -01 53 32.1     & $ 10.0$    &      &  0.1    &  9.9      &  0.5 \\
146      & H & 05 41 38.0   & -01 53 57.6     & $ 7.8$    &      &  0.1    & 7.7      & 0.4  \\
148      & H & 05 41 38.9   & -01 53 54.0     & $ 4.7$    &      &  0       & 4.7      & 0.2  \\
126      & H & 05 41 39.0   & -01 54 09.7     & $ 4.5$    &      &  0       & 4.5      & 0.2  \\
145      & H & 05 41 39.1   & -01 53 58.6     & $ 4.0$    &      &  0.1    & 3.9      & 0.2  \\
120      & H & 05 41 39.1   & -01 54 14.4     & $ 5.4$    &      &  0.1    & 5.3      & 0.3  \\
136      & H & 05 41 39.2   & -01 54 02.4     & $ 3.9$    &      &  0       & 3.9      & 0.2  \\
147      & H & 05 41 40.1   & -01 53 56.1     & $ 6.3$    &      &  0.1    & 6.2      & 0.3  \\
213      & I & 05 41 37.9   & -01 53 11.5     & $ 7.2$    &      &  0       & 7.2      & 0.4  \\
203      & I & 05 41 38.0   & -01 53 19.0     & $ 5.3$    &      &  0.1    & 5.2      & 0.3  \\
198      & I & 05 41 38.1   & -01 53 25.8     & $ 4.7$    &      &  0.1    & 4.6      & 0.2  \\
192      & I & 05 41 38.3   & -01 53 28.7     & $ 4.5$    &     &  0.1    & 4.4      & 0.2  \\
186      & I & 05 41 38.3   & -01 53 33.5     & $ 5.4$    &      &  0       & 5.4      & 0.3  \\
204      & I & 05 41 38.4   & -01 53 18.2     & $ 4.3$    &      &  0       & 4.3      & 0.2  \\
200      & I & 05 41 38.6   & -01 53 23.1     & $ 3.9$    &      &  0       & 3.9      & 0.2  \\
201      & I & 05 41 39.3   & -01 53 22.3     & $ 4.7$    &      &  0.1    & 4.6      & 0.2  \\
195      & I & 05 41 39.5   & -01 53 27.1     & $ 5.3$    &      &  0.1    & 5.2      & 0.3  \\
183      & I & 05 41 39.5   & -01 53 34.0     & $ 6.7$    &      &  0       & 6.7      & 0.4  \\
\enddata
\vskip -0.4in
\tablecomments{
Sources associated with asterisks were also detected at 3\,mm by 
\citet{eisner03}: IRC\,106/098 = Source 8, IRC\,232 = Source 9 (IRS\,2), 
IRC\,071 = Source 4, and IRC\,124 = Source 1.  Col.~(1): Source designation, 
according to \citet{meyerphd}.  Col.~(2): SMA field, as labeled in Figure 
\ref{pointings} and Table \ref{table-obs}.  Cols.~(3, 4): SMA emission centroid 
coordinates.  Col.~(5): Integrated continuum flux density, or 3$\sigma$ upper limits
on the non-detections, corrected for SMA 
primary beam attenuation.  Col.~(6): Extrapolated contribution of free-free 
emission at 887\,$\micron$ estimated from \citet{rodriguez03} measurements.
Col.~(7): Estimated contribution from large-scale cloud emission.  Col.~(8): 
Derived dust continuum flux density from the disk.  Col.~(9): Inferred disk 
mass (uncertainty does not include systematics in the absolute flux scale, 
which contribute an additional $\sim$10\%). 
}
\end{deluxetable}

\clearpage

\bibliographystyle{apj}
\bibliography{bib_rm}

\begin{thebibliography}{61}
\expandafter\ifx\csname natexlab\endcsname\relax\def\natexlab#1{#1}\fi

\bibitem[{{Adams} {et~al.}(2004){Adams}, {Hollenbach}, {Laughlin}, \&
  {Gorti}}]{adams04}
{Adams}, F.~C., {Hollenbach}, D., {Laughlin}, G., \& {Gorti}, U. 2004, \apj,
  611, 360

\bibitem[{{Akeson} \& {Jensen}(2014)}]{akeson14}
{Akeson}, R.~L., \& {Jensen}, E.~L.~N. 2014, \apj, 784, 62

\bibitem[{{Ali} {et~al.}(1998){Ali}, {Sellgren}, {Depoy}, {Carr}, {Gatley},
  {Merrill}, \& {Lada}}]{ali98}
{Ali}, B., {Sellgren}, K., {Depoy}, D.~L., {Carr}, J.~S., {Gatley}, I.,
  {Merrill}, K.~M., \& {Lada}, E. 1998, in Astronomical Society of the Pacific
  Conference Series, Vol. 154, Cool Stars, Stellar Systems, and the Sun, ed.
  R.~A. {Donahue} \& J.~A. {Bookbinder}, 1663

\bibitem[{{Andrews} {et~al.}(2013){Andrews}, {Rosenfeld}, {Kraus}, \&
  {Wilner}}]{andrews13}
{Andrews}, S.~M., {Rosenfeld}, K.~A., {Kraus}, A.~L., \& {Wilner}, D.~J. 2013,
  \apj, 771, 129

\bibitem[{{Andrews} \& {Williams}(2005)}]{andrews05}
{Andrews}, S.~M., \& {Williams}, J.~P. 2005, \apj, 631, 1134

\bibitem[{{Andrews} \& {Williams}(2007)}]{andrews07}
---. 2007, \apj, 671, 1800

\bibitem[{{Andrews} {et~al.}(2009){Andrews}, {Wilner}, {Hughes}, {Qi}, \&
  {Dullemond}}]{andrews09}
{Andrews}, S.~M., {Wilner}, D.~J., {Hughes}, A.~M., {Qi}, C., \& {Dullemond},
  C.~P. 2009, \apj, 700, 1502

\bibitem[{{Andrews} {et~al.}(2010){Andrews}, {Wilner}, {Hughes}, {Qi}, \&
  {Dullemond}}]{andrews10}
---. 2010, \apj, 723, 1241

\bibitem[{{Anthony-Twarog}(1982)}]{Anthony-twarog}
{Anthony-Twarog}, B.~J. 1982, \aj, 87, 1213

\bibitem[{{Baraffe} {et~al.}(1998){Baraffe}, {Chabrier}, {Allard}, \&
  {Hauschildt}}]{baraffe98}
{Baraffe}, I., {Chabrier}, G., {Allard}, F., \& {Hauschildt}, P.~H. 1998, \aap,
  337, 403

\bibitem[{{Barnes} {et~al.}(1989){Barnes}, {Crutcher}, {Bieging}, {Storey}, \&
  {Willner}}]{barnes}
{Barnes}, P.~J., {Crutcher}, R.~M., {Bieging}, J.~H., {Storey}, J.~W.~V., \&
  {Willner}, S.~P. 1989, \apj, 342, 883

\bibitem[{{Beckwith} {et~al.}(1990){Beckwith}, {Sargent}, {Chini}, \&
  {Guesten}}]{beckwith90}
{Beckwith}, S.~V.~W., {Sargent}, A.~I., {Chini}, R.~S., \& {Guesten}, R. 1990,
  \aj, 99, 924

\bibitem[{{Beuther} {et~al.}(2002){Beuther}, {Schilke}, {Menten}, {Motte},
  {Sridharan}, \& {Wyrowski}}]{beuther02}
{Beuther}, H., {Schilke}, P., {Menten}, K.~M., {Motte}, F., {Sridharan}, T.~K.,
  \& {Wyrowski}, F. 2002, \apj, 566, 945

\bibitem[{{Bik} {et~al.}(2003){Bik}, {Lenorzer}, {Kaper}, {Comer{\'o}n},
  {Waters}, {de Koter}, \& {Hanson}}]{bik}
{Bik}, A., {Lenorzer}, A., {Kaper}, L., {Comer{\'o}n}, F., {Waters},
  L.~B.~F.~M., {de Koter}, A., \& {Hanson}, M.~M. 2003, \aap, 404, 249

\bibitem[{{Bonnell} {et~al.}(2003){Bonnell}, {Bate}, \& {Vine}}]{bonnell03}
{Bonnell}, I.~A., {Bate}, M.~R., \& {Vine}, S.~G. 2003, \mnras, 343, 413

\bibitem[{{Da Rio} {et~al.}(2010){Da Rio}, {Robberto}, {Soderblom}, {Panagia},
  {Hillenbrand}, {Palla}, \& {Stassun}}]{dario}
{Da Rio}, N., {Robberto}, M., {Soderblom}, D.~R., {Panagia}, N., {Hillenbrand},
  L.~A., {Palla}, F., \& {Stassun}, K.~G. 2010, \apj, 722, 1092

\bibitem[{{D'Antona} \& {Mazzitelli}(1997)}]{dantona97}
{D'Antona}, F., \& {Mazzitelli}, I. 1997, \memsai, 68, 807

\bibitem[{{Di Francesco} {et~al.}(2008){Di Francesco}, {Johnstone}, {Kirk},
  {MacKenzie}, \& {Ledwosinska}}]{scuba}
{Di Francesco}, J., {Johnstone}, D., {Kirk}, H., {MacKenzie}, T., \&
  {Ledwosinska}, E. 2008, \apjs, 175, 277

\bibitem[{{Draine}(2006)}]{draine06}
{Draine}, B.~T. 2006, \apj, 636, 1114

\bibitem[{{Eisner}(2012)}]{eisner12}
{Eisner}, J.~A. 2012, \apj, 755, 23

\bibitem[{{Eisner} \& {Carpenter}(2003)}]{eisner03}
{Eisner}, J.~A., \& {Carpenter}, J.~M. 2003, \apj, 598, 1341

\bibitem[{{Eisner} \& {Carpenter}(2006)}]{eisner06}
---. 2006, \apj, 641, 1162

\bibitem[{{Eisner} {et~al.}(2008){Eisner}, {Plambeck}, {Carpenter}, {Corder},
  {Qi}, \& {Wilner}}]{eisner08}
{Eisner}, J.~A., {Plambeck}, R.~L., {Carpenter}, J.~M., {Corder}, S.~A., {Qi},
  C., \& {Wilner}, D. 2008, \apj, 683, 304

\bibitem[{{Evans} {et~al.}(2009){Evans}, {Dunham}, {J{\o}rgensen}, {Enoch},
  {Mer{\'{\i}}n}, {van Dishoeck}, {Alcal{\'a}}, {Myers}, {Stapelfeldt},
  {Huard}, {Allen}, {Harvey}, {van Kempen}, {Blake}, {Koerner}, {Mundy},
  {Padgett}, \& {Sargent}}]{evans09}
{Evans}, II, N.~J., {et~al.} 2009, \apjs, 181, 321

\bibitem[{{Feigelson} \& {Nelson}(1985)}]{feigelson}
{Feigelson}, E.~D., \& {Nelson}, P.~I. 1985, \apj, 293, 192

\bibitem[{{Grasdalen}(1974)}]{grasdalen74}
{Grasdalen}, G.~L. 1974, \apj, 193, 373

\bibitem[{{Haisch} {et~al.}(2000){Haisch}, {Lada}, \& {Lada}}]{haisch00}
{Haisch}, Jr., K.~E., {Lada}, E.~A., \& {Lada}, C.~J. 2000, \aj, 120, 1396

\bibitem[{{Haisch} {et~al.}(2001){Haisch}, {Lada}, {Pi{\~n}a}, {Telesco}, \&
  {Lada}}]{haisch01}
{Haisch}, Jr., K.~E., {Lada}, E.~A., {Pi{\~n}a}, R.~K., {Telesco}, C.~M., \&
  {Lada}, C.~J. 2001, \aj, 121, 1512

\bibitem[{{Harris} {et~al.}(2012){Harris}, {Andrews}, {Wilner}, \&
  {Kraus}}]{harris12}
{Harris}, R.~J., {Andrews}, S.~M., {Wilner}, D.~J., \& {Kraus}, A.~L. 2012,
  \apj, 751, 115

\bibitem[{{Henning} \& {Stognienko}(1996)}]{henning96}
{Henning}, T., \& {Stognienko}, R. 1996, \aap, 311, 291

\bibitem[{{Ho} {et~al.}(2004){Ho}, {Moran}, \& {Lo}}]{ho}
{Ho}, P.~T.~P., {Moran}, J.~M., \& {Lo}, K.~Y. 2004, \apjl, 616, L1

\bibitem[{{Johnstone} {et~al.}(1998){Johnstone}, {Hollenbach}, \&
  {Bally}}]{johnstone98}
{Johnstone}, D., {Hollenbach}, D., \& {Bally}, J. 1998, \apj, 499, 758

\bibitem[{{Lada}(1991)}]{lada91}
{Lada}, C.~J. 1991, in NATO ASIC Proc. 342: The Physics of Star Formation and
  Early Stellar Evolution, ed. {C.~J.~Lada \& N.~D.~Kylafis}, 329--+

\bibitem[{{Lada} \& {Lada}(2003)}]{lada03}
{Lada}, C.~J., \& {Lada}, E.~A. 2003, \araa, 41, 57

\bibitem[{{Levine} {et~al.}(2006){Levine}, {Steinhauer}, {Elston}, \&
  {Lada}}]{levine06}
{Levine}, J.~L., {Steinhauer}, A., {Elston}, R.~J., \& {Lada}, E.~A. 2006,
  \apj, 646, 1215

\bibitem[{{Mann} \& {Williams}(2009)}]{mann09b}
{Mann}, R.~K., \& {Williams}, J.~P. 2009, \apjl, 694, L36

\bibitem[{{Mann} \& {Williams}(2010)}]{mann10}
---. 2010, \apj, 725, 430

\bibitem[{{Mann} {et~al.}(2014){Mann}, {Di Francesco}, {Johnstone}, {Andrews},
  {Williams}, {Bally}, {Ricci}, {Hughes}, \& {Matthews}}]{mann14}
{Mann}, R.~K., {et~al.} 2014, \apj, 784, 82

\bibitem[{{Matsuyama} {et~al.}(2003){Matsuyama}, {Johnstone}, \&
  {Hartmann}}]{matsuyama}
{Matsuyama}, I., {Johnstone}, D., \& {Hartmann}, L. 2003, \apj, 582, 893

\bibitem[{{Meyer}(1996)}]{meyerphd}
{Meyer}, M.~R. 1996, PhD thesis, Max-Planck-Institut f{\"u}r Astronomie,
  K{\"o}nigstuhl 17, D-69117 Heidelberg, Germany
  <EMAIL>meyer@mpia-hd.mpg.de</EMAIL>

\bibitem[{{Meyer} {et~al.}(1997){Meyer}, {Calvet}, \& {Hillenbrand}}]{meyer97}
{Meyer}, M.~R., {Calvet}, N., \& {Hillenbrand}, L.~A. 1997, \aj, 114, 288

\bibitem[{{Meyer} {et~al.}(2008){Meyer}, {Flaherty}, {Levine}, {Lada},
  {Bowler}, \& {Kandori}}]{meyer08}
{Meyer}, M.~R., {Flaherty}, K., {Levine}, J.~L., {Lada}, E.~A., {Bowler},
  B.~P., \& {Kandori}, R. 2008, {Star Formation in NGC 2023, NGC 2024, and
  Southern L1630}, ed. B.~{Reipurth}, 662

\bibitem[{{Miyake} \& {Nakagawa}(1993)}]{miyake93}
{Miyake}, K., \& {Nakagawa}, Y. 1993, Icarus, 106, 20

\bibitem[{{Osterloh} \& {Beckwith}(1995)}]{osterloh}
{Osterloh}, M., \& {Beckwith}, S.~V.~W. 1995, \apj, 439, 288

\bibitem[{{Porras} {et~al.}(2003){Porras}, {Christopher}, {Allen}, {Di
  Francesco}, {Megeath}, \& {Myers}}]{porras}
{Porras}, A., {Christopher}, M., {Allen}, L., {Di Francesco}, J., {Megeath},
  S.~T., \& {Myers}, P.~C. 2003, \aj, 126, 1916

\bibitem[{{Reggiani} {et~al.}(2011){Reggiani}, {Robberto}, {Da Rio}, {Meyer},
  {Soderblom}, \& {Ricci}}]{reggiani11}
{Reggiani}, M., {Robberto}, M., {Da Rio}, N., {Meyer}, M.~R., {Soderblom},
  D.~R., \& {Ricci}, L. 2011, \aap, 534, A83

\bibitem[{{Ricci} {et~al.}(2011{\natexlab{a}}){Ricci}, {Mann}, {Testi},
  {Williams}, {Isella}, {Robberto}, {Natta}, \& {Brooks}}]{ricci11}
{Ricci}, L., {Mann}, R.~K., {Testi}, L., {Williams}, J.~P., {Isella}, A.,
  {Robberto}, M., {Natta}, A., \& {Brooks}, K.~J. 2011{\natexlab{a}}, \aap,
  525, A81

\bibitem[{{Ricci} {et~al.}(2010{\natexlab{a}}){Ricci}, {Testi}, {Natta}, \&
  {Brooks}}]{ricci10b}
{Ricci}, L., {Testi}, L., {Natta}, A., \& {Brooks}, K.~J. 2010{\natexlab{a}},
  \aap, 521, A66

\bibitem[{{Ricci} {et~al.}(2010{\natexlab{b}}){Ricci}, {Testi}, {Natta},
  {Neri}, {Cabrit}, \& {Herczeg}}]{ricci10}
{Ricci}, L., {Testi}, L., {Natta}, A., {Neri}, R., {Cabrit}, S., \& {Herczeg},
  G.~J. 2010{\natexlab{b}}, \aap, 512, A15+

\bibitem[{{Ricci} {et~al.}(2011{\natexlab{b}}){Ricci}, {Testi}, {Williams},
  {Mann}, \& {Birnstiel}}]{ricci11b}
{Ricci}, L., {Testi}, L., {Williams}, J.~P., {Mann}, R.~K., \& {Birnstiel}, T.
  2011{\natexlab{b}}, \apjl, 739, L8

\bibitem[{{Richling} \& {Yorke}(2000)}]{richling00}
{Richling}, S., \& {Yorke}, H.~W. 2000, \apj, 539, 258

\bibitem[{{Rodr{\'{\i}}guez} {et~al.}(2003){Rodr{\'{\i}}guez}, {G{\'o}mez}, \&
  {Reipurth}}]{rodriguez03}
{Rodr{\'{\i}}guez}, L.~F., {G{\'o}mez}, Y., \& {Reipurth}, B. 2003, \apj, 598,
  1100

\bibitem[{{Sault} {et~al.}(1995){Sault}, {Teuben}, \& {Wright}}]{sault}
{Sault}, R.~J., {Teuben}, P.~J., \& {Wright}, M.~C.~H. 1995, in Astronomical
  Society of the Pacific Conference Series, Vol.~77, Astronomical Data Analysis
  Software and Systems IV, ed. {R.~A.~Shaw, H.~E.~Payne, \& J.~J.~E.~Hayes},
  433--+

\bibitem[{{Scally} \& {Clarke}(2001)}]{scally01}
{Scally}, A., \& {Clarke}, C. 2001, \mnras, 325, 449

\bibitem[{{Sheehan} \& {Eisner}(2014)}]{sheehan14}
{Sheehan}, P.~D., \& {Eisner}, J.~A. 2014, \apj, 791, 19

\bibitem[{{Soderblom} {et~al.}(2013){Soderblom}, {Hillenbrand}, {Jeffries},
  {Mamajek}, \& {Naylor}}]{soderblom13}
{Soderblom}, D.~R., {Hillenbrand}, L.~A., {Jeffries}, R.~D., {Mamajek}, E.~E.,
  \& {Naylor}, T. 2013, ArXiv e-prints

\bibitem[{{Sridharan} {et~al.}(2002){Sridharan}, {Beuther}, {Schilke},
  {Menten}, \& {Wyrowski}}]{Sridharan02}
{Sridharan}, T.~K., {Beuther}, H., {Schilke}, P., {Menten}, K.~M., \&
  {Wyrowski}, F. 2002, \apj, 566, 931

\bibitem[{{St{\"o}rzer} \& {Hollenbach}(1999)}]{storzer}
{St{\"o}rzer}, H., \& {Hollenbach}, D. 1999, \apj, 515, 669

\bibitem[{{Williams}(2010)}]{williams10}
{Williams}, J. 2010, Contemporary Physics, 51, 381

\bibitem[{{Williams} {et~al.}(2005){Williams}, {Andrews}, \&
  {Wilner}}]{williams05}
{Williams}, J.~P., {Andrews}, S.~M., \& {Wilner}, D.~J. 2005, \apj, 634, 495

\bibitem[{{Williams} {et~al.}(2013){Williams}, {Cieza}, {Andrews}, {Coulson},
  {Barger}, {Casey}, {Chen}, {Cowie}, {Koss}, {Lee}, \& {Sanders}}]{williams13}
{Williams}, J.~P., {et~al.} 2013, \mnras, 435, 1671

\end{thebibliography}

\end{document}